\documentclass[useAMS,usenatbib]{mn2e}
\usepackage{epsfig}
\usepackage{color} 
\usepackage{natbib}
\usepackage[colorlinks=true,citecolor=blue]{hyperref}
\def\ew{$W_{r}$}
\def\mgi{Mg~{\sc i}~} 
\def\mgii{Mg~{\sc ii}~} 
\def\mgiia{Mg~{\sc ii}$\lambda$2796} 
\def\mgiib{Mg~{\sc ii}$\lambda$2803} 
\def\mgiiab{Mg~{\sc ii}$\lambda\lambda$2796,2803~} 
\def\feii{Fe~{\sc ii}~} 
 
\def\feiia{Fe~{\sc ii}$\lambda$2600~} 

\def\civ{C~{\sc iv}~}

\def\zem{$z_{\rm em}$~} 
\def\zemi{$z_{\rm em}$~} 
\def\lya{Ly$\alpha$~}

\def\chisq{$\chi^{2}$} 
\def\kms{km.s$^{-1}$} 
\newcommand{\rewmgiione}{\ensuremath{W_r^{\lambda2796}}}

\title[Strong {\mgii} Absorbers Towards Different Types of Quasars]{ 
Incidence of Strong {\mgii} Absorbers Towards Different Types of Quasars}
\author[$Joshi$, $Chand$ \& $Gopal-Krishna$ ]{Ravi Joshi$^{1}$\thanks{E-mail: ravi@aries.res.in(RJ);
hum@aries.res.in(HC); krishna@ncra.tifr.res.in(G-K)}, Hum Chand$^{1}$$^{\star}$ ,
Gopal-Krishna$^{2}$$^{\star}$ \\ $^{1}$Aryabhatta Research Institute of Observational Sciences (ARIES),
Manora Peak, Nainital $-$ 263129, India\\\\ $^{2}$Inter-University
Centre for Astronomy and Astrophysics (IUCAA), Postbag 4, Ganeshkhind, Pune 411 007, India\\}

\begin{document}
\date{Accepted ---. Received ---; in original form ---}

\pagerange{\pageref{firstpage}--\pageref{lastpage}} \pubyear{2009}

\maketitle

\label{firstpage}
\begin{abstract}
We report the first comparative study of strong \mgii absorbers (\ew
$\ge 1.0$\AA) seen towards radio-loud quasars (RLQs) of 
core-dominated (CDQ) and lobe-dominated (LDQ) types and normal quasars
(QSOs). The CDQ and LDQ samples were derived from the Sloan Digital
Sky Survey data release 7 (SDSS-DR7) after excluding known
`broad-absorption-line' quasars (BALQSOs) and blazars. The \mgii
associated absorption systems having a velocity offset $v < 5000$~\kms~ from the
systemic velocity of the background quasar were also excluded.
Existing spectroscopic data for redshift-matched sightlines of 3975
CDQs and 1583 LDQs, covering a emission redshift range 0.39$-$4.87,
were analysed and 864 strong \mgii absorbers were found,
covering the redshift range 0.45$-$2.17. The conclusions reached using
this well-defined large dataset of strong \mgii absorbers are: (i) The
number density, $dN/dz$, towards CDQs shows a small, marginally
  significant excess ($\sim 9\%$ at 1.5$\sigma$ significance) over
the estimate available for QSOs; (ii) In the redshift space, this
difference is reflected in terms of  a $1.6\sigma$ excess of $dN/dz$ over the
QSOs, within the narrow redshift interval 1.2 $-$ 1.8; (iii) The
  $dN/d\beta$ distribution (with $\beta = v/c$) for CDQs shows a
  significant excess (at $3.75\sigma$ level) over the distribution
  found for a redshift and luminosity matched sample of QSOs, at
  $\beta$ in the range 0.05$-$0.1.  This leads us to infer that
a significant fraction of strong \mgii absorption systems seen in this
offset velocity range are probably associated with the CDQs and might
be accelerated into the line of sight by their powerful jets and/or
due to the accretion-disk outflows close to our direction. Support to
this scenario comes from a consistency check in which we only
consider the spectral range corresponding to $\beta$~$>$ 0.2. The
computed redshift distribution for strong \mgii absorbers towards CDQs
now shows excellent agreement with that known for QSOs, as indeed is
expected for purely intervening absorption systems. Thus, it appears
that for CDQs (and blazars) the associated strong \mgii absorbers can
 be seen at much larger velocities relative to the nucleus than the commonly
adopted upper limit of 5000~\kms.
\end{abstract}


\section{Introduction}
\label{sec:intro_mgiidndz}

The analysis of quasar absorption line systems (QALs) has emerged
as a powerful means of investigating the physical conditions of the
gaseous medium of the intervening galaxies also when they lie at very
large redshifts and are too faint for direct imaging/spectroscopy even
with the largest telescopes \citep{Bahcall1966ApJ...144..847B,
  Bahcall1980srst.coll..215B, Wolfe2005ARA&A..43..861W, Kulkarni2012ApJ...749..176K}. Further,
recent studies have revealed the potential of QALs as tracers of
extremely high-speed (mildly relativistic) ejection of cool gas clouds
by the powerful jets emanating from radio-loud quasars (\citealt*{
  Bergeron2011A&A...525A..51B}, henceforth BBM;
\citealt{Chand2012ApJ...754...38C}, Paper I). Among the best studied
population of intervening galaxies is the one selected by \mgii
absorption systems, which based on their rest-frame equivalent width
(\ew), whether greater than 1\AA\ or not, are classified as
`strong systems' and `weak systems', respectively. It is now well
established that for strong \mgii system (\ew $\ge 1.0$\AA) a galaxy
is nearly always found within an impact parameter of $\sim$ 100 kpc
\citep[e.g.,][]{Steidel1995qal..conf..139S,
  Churchill2005pgqa.conf...24C, Zibetti2007ApJ...658..161Z,
  Kacprzak2008AJ....135..922K, Chen2008ApJ...687..745C} and spanning a
wide range in optical luminosity \citep[e.g.,][]
{Bergeron1986A&A...169....1B, Steidel1995qal..conf..139S,
  Kacprzak2008AJ....135..922K}. On the other hand, weak \mgii system
($0.3$\AA\ $\le$ \ew $ < 1.0$\AA) are thought to predominantly trace
separate populations of galaxies, such as low surface brightness
galaxies, or dwarf galaxies \citep[e.g.,][]
{Churchill1999ApJS..120...51C, Narayanan2007ApJ...660.1093N}.

  Until recently, the paradigm has been that the cool gas clouds
  (e.g., \mgii absorption systems) with velocities $\beta c \la 5000$
  \kms\ relative to the background quasar are mostly associated with
  the quasar itself (`associated systems'), although somewhat higher
  velocity limit has been inferred in recent literature
  \citep[e.g.,][]{Wild2008MNRAS.388..227W}. Absorbers at significantly
  larger velocity offsets are believed to be `intervening' systems
  entirely unrelated to the background quasar. Recent evidence,
  however, appears to question this canonical view and suggests that
  even associated systems can have significantly relativistic
  velocities relative to the background quasar, in case the quasar is
  undergoing powerful jet activity and/or high speed accretion-disk
  outflows (see below). A possible signature of this would be if the
  occurrence rate, $dN/dz$, of `intervening' absorbers is found to
  depend on the type of the background source. Indeed, such an
  indication came from the unexpected finding of BBM that $dN/dz$ of
  \mgii absorption systems towards blazars is $\sim$ 2 times (at
  3$\sigma$ confidence) higher than the value known for normal
  (optically selected and mostly radio-quiet) quasars (QSOs), when
  strong \mgii lines alone are considered. A similar excess of
  $\sim1.8$ times (at 2.5$\sigma$ confidence) was also seen for the weak
  \mgii absorption systems by BBM. However, this result
  for blazars, which are as a class, believed to have relativistic
  jets pointed close to our direction, is drawn from a modest size
  sample consisting of only 45 powerful blazars having $0.8 < z_{em} <
  1.9$.
On the other hand, no significant excess of
$dN/dz$ was found by \citet{Chand2012ApJ...754...38C} in their
analysis of the spectral data for radio core-dominated quasars (CDQs,
also called flat-spectrum radio quasars: FSRQs), whose flux is also jet
dominated, albeit only weakly polarized.

These authors also presented evidence that the observed excess of
\mgii absorption systems towards blazars is confined to the velocity
offsets of up to $\sim$ 0.15c, relative to the background source.
Thus, they proposed that the excess of \mgii absorbing clouds seen
towards blazars is not evident towards CDQs because their jets are
less closely aligned with our direction than are the blazar jets and,
consequently, the putative absorbing clouds accelerated to mildly
  relativistic speeds by those jets are unlikely to intersect the
line-of-sight. This explanation conforms to the orientation based
unified model for powerful radio sources, according to which powerful
blazars, core-dominated quasars (CDQs), lobe-dominated quasars (LDQs)
and classical radio galaxies (RGs) form an orientation sequence
characterized by an increasing degree of the jet's mis-alignment
  from the line of sight \citep[as reviewed, e.g.,
  by][]{Urry1995PASP..107..803U, Barthel1999ASPC..162..127B,
  Antonucci2012A&AT...27..557A}. \par

In order to further probe the role of the quasar jets in accelerating
the absorption-line clouds to extremely large velocities, we endeavor
here to extend the investigation of strong \mgii absorption systems by
one more step along the orientation sequence, i.e., by including LDQs.
To do this, we shall examine multiple attributes of \mgii absorption
in uniformly selected, well-matched large samples of CDQs and LDQs and
compare the $dN/dz$ and $dN/d\beta$ for these two quasar types, and
also with the corresponding estimates available for optically selected
`normal' quasars (QSOs) which are known to be mostly radio-quiet and
lack powerful jets. As
explained below, the definitions of CDQs and LDQs used here, although
somewhat unconventional, basically conform to the intended difference
in terms of mis-alignment of the radio jet from the line-of-sight, as
mentioned above. \par

We have taken the CDQ/LDQ classification for each quasar from the SDSS
catalog which itself is based on the criterion adopted in
\citet{Jiang2007ApJ...656..680J}. Although seemingly different, their
criterion is effectively consistent with the standard definition based
on either radio spectral index, or the fraction of total radio flux
contributed by the nuclear core. In the definition of
\citet{Jiang2007ApJ...656..680J} a quasar is CDQ if its radio
counterpart in the FIRST survey (resolution 5 arcsec,
\citealt{Becker1995ApJ...450..559B}) is either a single component, or
is unresolved, or only partially resolved. On the other hand, a
multi-component radio counterpart in the FIRST survey (within a 30
arcsec region) is designated by them as LDQ. Despite being
unconventional, such a classification scheme meets the basic
requirement for the purpose of the present statistical study, since
the axes of the quasars classified by them as LDQs can be expected to
have, on average, a larger mis-alignment from our direction, in
comparison to the quasars classified by them as CDQs. Therefore, we
have adopted here the CDQ/LDQ classification as provided in the
SDSS-DR7 catalog. \par

  Here, we may recall the few alternative physical explanations that
have been considered for the reported excess of $dN/dz$ 
  towards blazars vis a vis QSOs. These are (i) obscuration of the
background QSO by the foreground dust, possibly associated with the
\mgii absorbers themselves, and (ii) gravitational lensing of 
the background blazar by intervening galaxies, possibly the ones hosting the \mgii
absorbers \citep[see also,][]{frank2007Ap&SS.312..325F,
  Hao2007ApJ...659L..99H}. Due to the dust obscuration, one might miss
the QSOs (especially, the optically-selected QSOs); however, the
expected change in $dN/dz$ of \mgii systems due to these effects is
estimated to be minute (only $\sim$ 1\%)
\citep[e.g.][]{Menard2008MNRAS.385.1053M}. BBM have argued that the
expected amplitude of dust obscuration, or gravitational lensing falls
short, by at least an order of magnitude, of explaining the
afore-mentioned factor of two excess in the incidence of \mgii
absorbers towards blazars, vis a vis normal QSOs. As an alternative
possibility, they have argued that the powerful blazar jets are
capable of sweeping cool gas clouds of sufficiently large column
densities (up to $10^{18} - 10^{20}~cm^{-2}$) and accelerating them to
velocities of order $0.1c$, potentially accounting for the observed
excess of \mgii absorption systems towards blazars, as compared to
normal QSOs. \par

This paper is organised as follows. Sections 2 and 3 describe the
selection of the samples of CDQs and LDQs, as well as the procedure
for analysis of their optical spectra. In Section 4, we present the
results of our analysis, followed by a discussion in Section 5.

\section{The CDQ and LDQ samples}
\label{sec:sample_mgiidndz}

Our samples of CDQs and LDQs are extracted from the quasar catalog of
\citet{shen2011ApJS..194...45S} (their online Table 1) which is
derived from the Sloan Digital Sky survey, Data Release 7 (SDSS-DR7;
\citep{Abazajian2009ApJS..182..543A, Schneider2010yCat.7260....0S}. We
have applied five selection filters. Firstly, the radio-loudness
parameter, R (= ratio of flux densities at 5 GHz and at 2500\AA\ in
the rest-frame) must exceed 10 \citep{Kellermann1989AJ.....98.1195K}.
This resulted in 8257 radio-loud sources. The next filter was the
availability of radio classification (CDQ, or LDQ), which was in fact
available for all 8257 quasars in the \citet{shen2011ApJS..194...45S}
compilation\footnote{0=FIRST undetected; 1=core-dominated;
  2=lobe-dominated (radio morphology classification following
  \citealt{Jiang2007ApJ...656..680J})}, based on the FIRST survey, as
already explained in Section~\ref{sec:intro_mgiidndz}. This filter
resulted in 6152 CDQs and 2105 LDQs (the latter having a FIRST radio image
resolved into multiple components). A third filter was applied to
exclude the sources with \zemi smaller than 0.389, for which \mgii
emission line falls below the SDSS wavelength coverage
(3800\AA-9200\AA in the observer's frame of reference). This reduced
the sample to 5824 CDQs and 2014 LDQs. The fourth filter was then
applied by selecting only non-BAL quasars, which are flagged in the
\citet{shen2011ApJS..194...45S} catalog, resulting in 5447 CDQs and
1971 LDQs. The final filter excluded all sources marked in the
\citet{veron2010A&A...518A..10V} catalog as `HP' (i.e., high optical
fractional polarization), or BL Lac, or Seyfert. This selection chain
led to a final sample containing 5333 CDQs and 1925 LDQs (total 7258
radio-loud quasars).

\begin{figure}
 \epsfig{figure=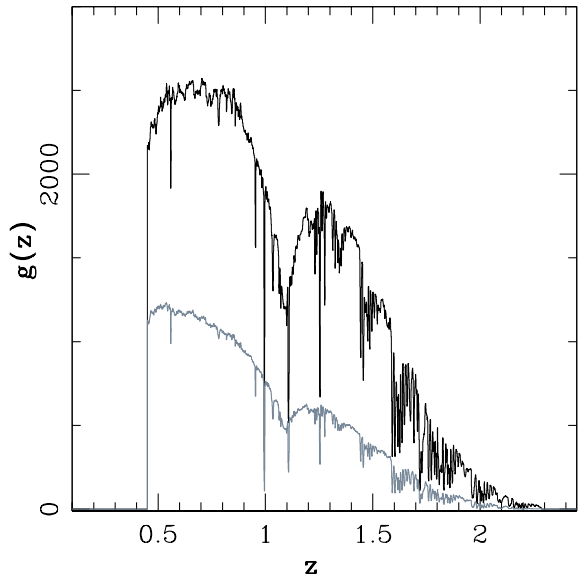,height=8.cm,width=8.cm,angle=0}
  \caption{Redshift path density, $g(z)$, for the strong (\ew $\ge$ 1.0\AA)
    `intervening' \mgii systems detected towards our CDQ (black curve) and 
    LDQ (grey curve) samples.}
 \label{fig:goz}
 \end{figure}

The reduced one-dimensional (1D) spectra for our entire sample of 7258
quasars were downloaded from the SDSS Data Archive
Server\footnote{mirror.sdss3.org/bulk/Spectra} (detail about the SDSS
spectral information can be found
in\citealt{York2000AJ....120.1579Y}). Briefly, the SDSS spectra cover
a spectral range from 3800\AA ~\ to 9200\AA, with a resolution
($\lambda/\Delta \lambda$) of about 2000 (i.e., 150~\kms). This
spectral coverage allowed us to search for \mgii absorption systems
over a redshift range 0.389 $<~z_{abs}~<$ 2.27. The median redshifts
of the redshift paths covered by our CDQ and LDQ samples are essentially
identical, being 1.37 and 1.36, respectively (see
Figure~\ref{fig:goz}). This is not surprising since the spectra of
both correspond to the SDSS spectral coverage and hence have the same
observed wavelength window. This redshift matching is a well known
essential requirement for the purpose of comparing $dN/dz$ of
intervening \mgii absorbers in quasar samples. Equally important is to
ensure a matching range of $z_{em}$, before comparing the
distributions of the relative velocity ($v$) of the absorption systems
relative to the respective background sources of different types
(Section~\ref{subsec:beta_intorigin}). At the same time, for the relative
  velocity distribution to be free from any systematic bias, accuracy
  of \zemi measurements is a crucial requirement.
  Recently,~\citet{Hewett2010yCat..74052302H} have refined the SDSS
  emission redshift values by reducing the net systematic errors by
  almost a factor of 20, attaining an accuracy of up to $\sim$30\kms.
  Therefore, throughout our analysis we have used the emission
  redshift values for our entire sample taken from this new catalog by
  \citet{Hewett2010yCat..74052302H}.  \par

 \begin{figure*}
\epsfig{figure=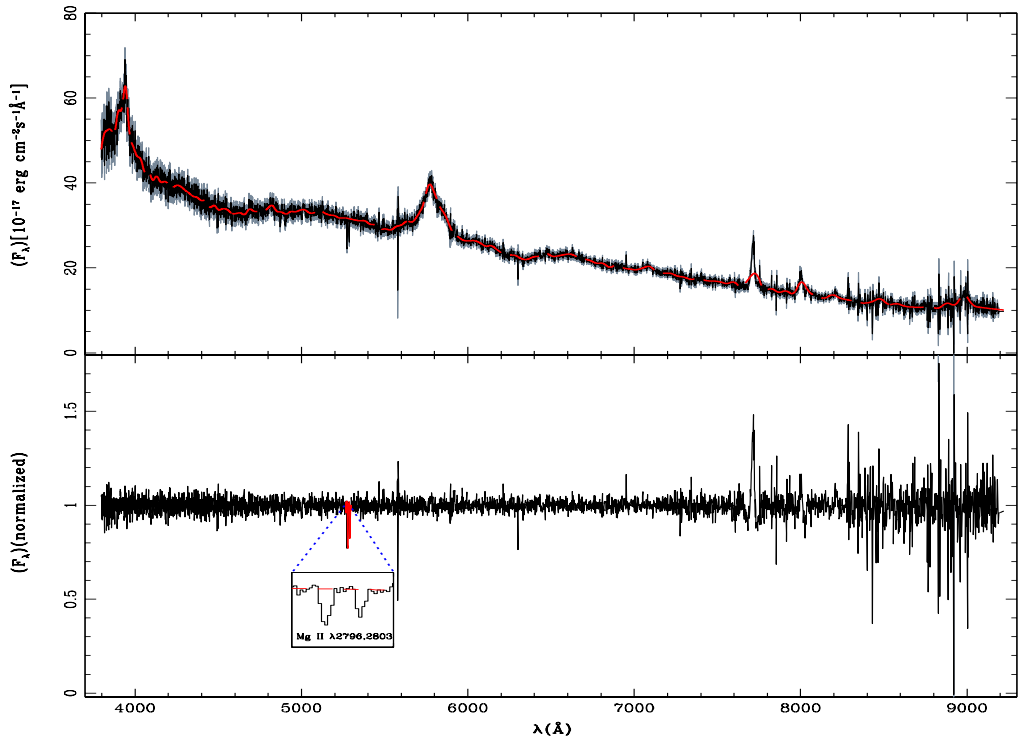,height=10.cm,width=14.cm,angle=00,bbllx=29bp,bblly=172bp,bburx=573bp,bbury=701bp,clip=true}
  \caption{ \emph{upper panel:} SDSS spectrum of the quasar 
J002745.23$-$094603.3(black), the corresponding 1$\sigma$ error bar range (grey) 
and the continuum fit (red). The \emph{lower panel} shows the normalized 
spectrum, the inset displays the zoom-in on the \mgii absorption doublet.}
 \label{fig:continuum_fit}
 \end{figure*}

\begin{table}
\centering
\begin{minipage}{180mm}
\caption{Summary of the step-wise selection and the criteria used.}
\label{tab:dndz_abs_sumry}
\begin{tabular}{@{}llrr@{}}
\hline 

\multicolumn{1}{l}{}&\multicolumn{1}{l}{Selection criteria}
&\multicolumn{1}{r}{CDQs} & \multicolumn{1}{r}{LDQs}  \\
\hline 
\\
\multicolumn{2}{c}{For sample selection } & \multicolumn{2}{c}{Number of CDQs/LDQs } \\
\\
                 1.      &     Radio loudness ($R>10$)             & 6152 & 2105 \\
                 2.     &     \zem $>$ 0.389                       & 5824 & 2014 \\
                 3.      &     after excluding BALQSOs             & 5447 & 1971 \\
                 4.      &     after excluding  HP, BLlac, Sy      & 5333 & 1925 \\
                 5.      &     $SNR > 5$ (over 121-pixels)         & 3975 & 1583 \\
\\
\multicolumn{2}{c}{For \mgii Doublet identification   } & \multicolumn{2}{c}{Number of absorbers }\\
\\
                 6.     &   \mgiiab candidates            & 66726 & 16926\\
                 7.     &     doublet with \ew $>$ 0.5\AA (\mgiia)   & 15096 & 4316\\
                        &     and \ew $>$0.2\AA(\mgiib)              &       &  \\
                 8.     &     visual confirmation                    & 1113  &  424 \\
                 9.     &    $SNR > 5$ (over 121-pixels) and \ew $>$1\AA   & 640   &  224\\

\hline      
                                                      
\end{tabular} 
\end{minipage}
\end{table}

\section{Analysis}
\label{sec:analysis_mgiidndz}

\subsection{Identification of \mgii absorption systems}
\label{subsec:mgii_iden_mgiidndz}

For each quasar, identification of the \mgii absorption doublet in the
normalized continuum spectrum, was carried out using several publicly
available routines and, in particular, following the procedure
outlined in \citet{prochter2006ApJ...639..766P}\footnote{e.g.
  cvs.ucolick.org/viewcvs.cgi/xmen/xidl/SDSS/}. Basically, this
algorithm initially fits a continuum to the SDSS spectroscopic data,
employing first a principal component analysis (PCA) as a guess
for the \lya and \civ emission lines (i.e., from 1000\AA -2000\AA~ in
the rest-frame). Next, a b-spline algorithm is used to fit the
underlying residual continuum, which results roughly in a power-law
with broad emission features superposed. An example is shown in
Figure~\ref{fig:continuum_fit}. \par

This procedure automatically searches for absorption features in the
normalized spectrum redward of \lya. The search was carried out for
absorption features, which are fitted with Gaussian profile by taking
an initial FWHM of 2.5 pixels, with the additional requirement that
the minimum separation between lines should be about two times the
FWHM. Out of all such cases, the final selection was made by accepting
only the lines which are above two times the rms($\sigma$) noise in
the spectrum. \par

The absorption features thus identified in each quasar spectrum, were
searched for absorption line pairs. For this purpose our procedure
first computed redshift of a given absorption feature, assuming it to
be \mgiia. The corresponding positions of the expected \mgiib\ and
\feiia lines were then inspected. The criterion for accepting a
feature as genuine \mgii absorption system was that at least two of
these three lines must be present at the expected locations above a
2$\sigma$ threshold. Equivalent widths of all the accepted \mgii
absorption lines were then measured (in the observed frame), by
summing the difference from unity of the flux in the normalized
observed-frame spectrum, within about 10 pixels wide boxes (11.51\AA~
in the observed frame) placed at the centroids of the two \mgii lines.
In the spectra of 5333 CDQs and 1925 LDQs, our automated routine has
detected 66726, 16926 possible \mgii doublet candidates, out of which
we have selected only those candidates having their \ew~value for
\mgiia$\ge 0.5$\AA\ and \mgiib $\ge 0.2$\AA. This selection filter
resulted in 15096 and 4316 possible strong(\ew $> 1.0\AA$) \mgii
doublet candidates for CDQ and LDQ samples, respectively. \par

As a further check, we also carried out a visual confirmation of each
absorption system identified via the above automated procedure. This
step is important since (i) our automated search does not carry out
the line profile matching of the \mgii candidate doublet and hence can
result in over counting the \mgii doublets rather than missing out any
genuine systems, and (ii) any visually noticed uncertainty in the
continuum level could have significantly distorted the estimate of
\ew, rendering the strong/weak classification of absorption systems
unreliable. In this process of visual scrutiny, we first looked for
the strongest five \feii lines corresponding to the candidate \mgii
doublet. We then made a velocity plot of the \mgi, \mgii and the five
\feii lines, so as to match by eye the line profiles and their
strengths to those expected on the basis of the line oscillator
strengths. Thus, in the spectra of the 5333 CDQs, we visually
inspected all 15096 \mgii absorption systems candidates and confirmed
1113 of them. Likewise, a total of 424 \mgii absorption system were
confirmed out of the 4316 candidates visually inspected in the spectra
of the 1925 LDQs. Continuum fitting of each confirmed \mgii absorption
system was then further checked by plotting the fitted continuum to
the spectral segment containing the \mgii doublet. In all cases where
the continuum fitting over the relevant spectral segment seemed
unsatisfactory, we refined the local continuum fit and recomputed the
\ew({Mg~{\sc ii}}). Summary of our step-wise selection and the
criteria used to make our final sample are given in Table~\ref{tab:dndz_abs_sumry}. \par

\subsection{Number density of `intervening' strong \mgii absorption systems}

\label{subsec:number_den_mgii_dndz}

In order to determine the number density, $dN/dz$, of \mgii absorption
systems, we first excluded the redshift segments unsuited for a secure
detection of the \mgii absorption lines: (i) the regions blueward of
the Ly-$\alpha$ emission line, as it is strongly contaminated by the
Ly-$\alpha$ forest; (ii) the spectral regions within $5000$~\kms\ of
the \mgii emission line for a given background quasar, as any such
absorbing clouds are generally thought to have a high probability of
being associated with the quasar; (iii) the spectral regions polluted
by the various known atmospheric absorption features. This was
somewhat complicated, since some spectra were more affected by
telluric lines than the others. Simply excluding these regions equally
from each quasar spectrum would have led to an unnecessarily excessive
exclusion out of the available redshift path-length coverage.
Therefore, we have only excluded the obvious cases of blending
revealed by a visual inspection of each \mgii system candidate picked
by our automated algorithm. In addition, the uncertainty arising from
very noisy spectral segments was minimized by demanding that the
average SNR per pixel averaged over a 121-pixel box centered on the
pixel in question, be $>$ 5 and the member of the stronger component
of the \mgii absorption doublet is detectable above the designated \ew
threshold (which is 1.0\AA\ for strong systems) at more than
$3\sigma_{det}(\lambda)$ level ~\citep[e.g see
  also,][]{Lawther2012A&A...546A..67L}. Here, $\sigma_{det}(\lambda)$
is determined by integrating the noise spectrum over a resolution
element($\lambda/R$) around each pixel. For the 1358 spectra in our
CDQ sample and 342 spectra in our LDQ sample, the SNR in their entire
spectral range was found to be below this desired SNR threshold of 5
(averaged over a 121-pixel box) and hence they were dropped out from
our subsequent analysis. This leads to a final sample of 3975 CDQs and
1583 LDQs having 640 and 224 strong \mgii absorbers, respectively.

\par

We have taken the form of redshift path density $g(W_{min}, z)$, which
at any given redshift is the path in the redshift space available to
search for \mgii systems in our data sample, as

\begin{equation}
g(W_{min},z)=\Sigma_j H(z - z_j^{min}) H(z_j^{max} - z) 
H[W_{min}-W^j_{min}(z)]
\end{equation}

Where, $H$ is the Heaviside step function, $z^{min,max}$ correspond to
the minimum and maximum \mgii absorption redshifts allowed by the
spectral range considered for our search of \mgii doublet and
$W^j_{min}(z)$ is lower detection limit on \ew at
$3\sigma_{det}(\lambda)$ level allowed by the SNR per resolution
element of the spectrum. In Figure~\ref{fig:goz} we plot
$g(W_{min}=1.0$\AA$,z)$ available at minimum of
$3\sigma_{det}(\lambda)$ detection significance, for our sample of
3975 CDQs and 1583 LDQs. The redshift path, $\Delta z$, for our sample
between redshift $z^{min,max}$ is then given by

\begin{equation}
\Delta z = \int_{z^{min}}^{z^{max}} g(W_{min},z) dz
\label{eq:deltaz}
\end{equation}

The number density of absorbers per unit redshift is $dN/dz
=N_{obs}/\Delta z$, where $N_{obs}$ is the number of the \mgii
absorption systems with $W \ge W_{min}$, detected over the redshift
path $\Delta z$ at a significance $\ge 3\sigma$ level. Using this
formalism, we find for our sample of 3975 CDQs:

\begin{equation}
\frac{dN}{dz} (W_{\rm r}(2796)>1.0\,\AA)=\frac{N_{obs}(CDQ)}{\Delta z
  (CDQ)}=\frac{640.00}{2333.01}= 0.27^{+ 0.01}_{- 0.01
\label{eq:dndz_core}
}
\end{equation}

\noindent with mean value of the redshift path being $\langle \Delta z
\rangle\simeq 1.01$. For the sightlines towards our sample of 1583
LDQs, we similarly find:

\begin{equation}
\frac{dN}{dz} (W_{\rm r}(2796)>1.0\,\AA)=\frac{N_{obs}(LDQ)}{\Delta  z(LDQ)}=\frac{ 224.00}{ 922.11}= 0.24^{+ 0.02}_{- 0.02
\label{eq:dndz_lobe}
} 
\end{equation}

\noindent with a mean value of the redshift path being $\langle \Delta
z \rangle\simeq 0.93$. Here, the errors on $dN/dz$ correspond to
1$\sigma$ confidence level, and are computed using the standard
Poisson statistics. However, for $N_{obs} < 100$ (small number
statistics), e.g., in the subsequent analysis
(Section~\ref{sec:results_mgiidndz}), errors will be computed using
the limits corresponding to 1$\sigma$ confidence level of a Gaussian
distribution, as tabulated by \citet{Gehrels1986ApJ...303..336G}.
 Using Eq.~\ref{eq:dndz_core} and Eq.~\ref{eq:dndz_lobe} one can
compute the  ratio (excess factor) of $dN/dz$ values for strong \mgii absorber
between CDQ and LDQ sightlines, as
       
\begin{equation}
 Ex = \left( \frac{dN}{dz} \right)_{\rm CDQs} / \left( \frac {dN}{dz}
 \right)_{\rm LDQ} = 1.13\pm^{ 0.10}_{ 0.10},
 \label{eq:ex_cdq}
 \end{equation}

 which is found to differ only at a 1.3$\sigma$ level. \par

\begin{figure}    
   \centering    
      \includegraphics[width=7.0cm]{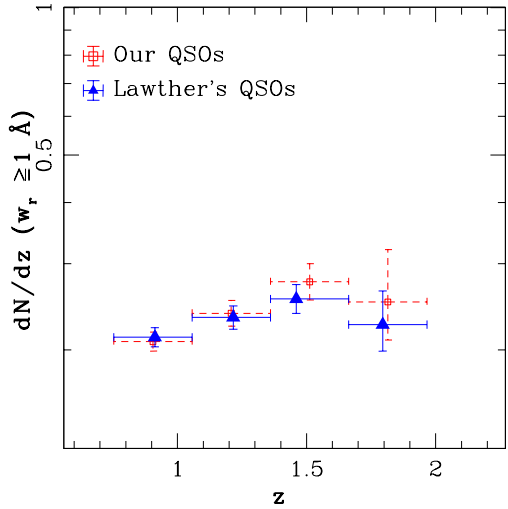}     
\caption{Redshift dependence  of the number density of
  intervening strong ($W_{\rm r}(2796) \ge 1.0$~\AA) \mgii absorption
  systems, based on the 3366 SDSS normal quasars common between our 
 sample (squares) and  \citet{Lawther2012A&A...546A..67L}  catalog (triangles), showing  a 
 very good   statistical agreement among them with a \chisq-test $P_{null}=0.89$ .}
         \label{fig:dNdz_lawther_n_ours}
   \end{figure}

 As a check on our analysis procedure, we have tested our method
   of $dN/dz$ estimation with the recent result reported in
   \citet{Lawther2012A&A...546A..67L} catalog, using their redshift path
 information for each sightline (priv. comm. with Dr. D. Lawther). For this we have
   derived a subset of 3366 normal quasars from the QSO catalog of
   \citet{Lawther2012A&A...546A..67L}, members of which are closely matched in emission
   redshift and bolometric luminosity with members of our CDQ sample. By
   downloading their spectra from the SDSS, we have computed $dN/dz$
   for \mgii strong absorbers, following the analysis procedure
   described above and then compared it with that obtained using
   the results provided in \citet{Lawther2012A&A...546A..67L} catalog.
   A very good statistical agreement is found between the two
   estimates of $dN/dz$, with a \chisq-test giving $P_{null} = 0.89$
   (e.g., see Figure ~\ref{fig:dNdz_lawther_n_ours}), validating our
   analysis procedure.

   \begin{figure}	
   \centering	
      \includegraphics[width=7.0cm]{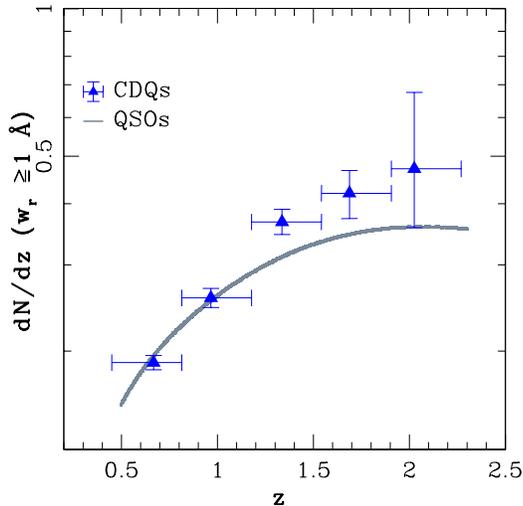}
\caption{ Redshift evolution of the number density of
  intervening strong ($W_{\rm r}(2796) \ge 1.0$~\AA) \mgii absorption
  systems seen towards our matched set of CDQs (triangles) and QSOs
  (solid curve). The solid curve is computed from the analytical
  expression given by \citet{Zhu2013ApJ...770..130Z} for strong \mgii
  absorption systems. The spectral region with offset velocities up to
  $5000$~\kms\ has been excluded from the analysis
  (Section~\ref{subsec:number_den_mgii_dndz}).}
         \label{fig:dNdz}
   \end{figure}

\section{Results}
\label{sec:results_mgiidndz}

\subsection{Comparison of $dN/dz$ for CDQs and LDQs versus normal QSOs}
\label{subsec:comp_dndz_CLQ}

We now compare the incidence rate of strong \mgii absorbers found here
along the CDQ and LDQ sightlines with that known for the sightlines
towards normal quasars (QSOs). For this, we have computed the number
per unit redshift towards QSOs, using the analytical fit recently
published by \citet{Zhu2013ApJ...770..130Z}. The fit is based on their
analysis of the \mgii absorbers towards QSOs in the SDSS-DR7, where
they have defined the incidence rate of \mgii\ absorbers in a given
range of rest-frame equivalent width \ew\ as:

\begin{eqnarray}
\frac{dN}{dz} (W_{\mathrm{min}}<W_r<W_{\mathrm{max}}) &=&
\int^{W_\mathrm{max}}_{W_\mathrm{min}} \frac{\partial^2 N}{\partial z \partial W_r} d W_r~.
\label{eq:fit_zhu}
\end{eqnarray}

\noindent where
\begin{equation}
\frac{\partial^2 N}{\partial z \partial \rewmgiione} (\rewmgiione, z)
= s(z)~e^{-\frac{\rewmgiione}{W^{\star}(z)}} \mathrm{,}
\end{equation}
\begin{equation}
s(z) = s_0~\frac{(1+z)^{\gamma_s}}{1+(\frac{z}{z_s})^{\delta_s}}
\mathrm{,}\nonumber
\end{equation}
\noindent and
\begin{equation}
W^{\star}(z) = W^{\star}_r~\frac{(1+z)^{\gamma_W}}{1+(\frac{z}{z_W})^{\delta_W}}
\mathrm{,}\nonumber
\end{equation}

\noindent in which $\gamma_{s}$, $\gamma_{W}$, $\delta_{s}$,
$\delta_{W}$, $z_{s}$, and $z_{W}>0$ with their best-fit values are as
listed in their Table~2. The values we have used here for
$W_{\mathrm{min}}$, $W_{\mathrm{max}}$ are 1.0\AA~ and 5.0\AA~,
respectively.

  To begin with, we computed the expected number of strong \mgii
  absorbers for the QSO sightlines ($N_{qso}$) corresponding to the
  same redshift path as available for the CDQs/LDQs sightlines, by
  multiplying Eq.~\ref{eq:fit_zhu} with the total redshift path for
  the CDQs/LDQs. These are found to be N$_{\rm exp}(QSO) =587.26$ and
  $221.58$, corresponding to the total redshift paths towards the CDQs and
  LDQs, respectively.  As a result the  expected $dN/dz$  towards normal QSOs
  corresponding to the total redshift paths towards the CDQs and
  LDQs are found, respectively, as 

\begin{equation}
\left( \frac{dN}{dz} \right)_{QSO} =\frac{N_{exp}(QSO)}{\Delta z
  (CDQ)}=\frac{587.26}{2333.01}= 0.25^{+ 0.01}_{- 0.01
\label{eq:dndz_core_qso}
}
\end{equation}

\begin{equation}
\left( \frac{dN}{dz} \right)_{QSO} =\frac{N_{exp}(QSO)}{\Delta z
  (LDQ)}=\frac{221.58}{922.11}= 0.24^{+ 0.02}_{- 0.02
\label{eq:dndz_lobe_qso}
}
\end{equation}

  These expected $dN/dz$  towards normal QSOs along with the use of 
  Eq.~\ref{eq:dndz_core} and Eq.~\ref{eq:dndz_lobe}  for $dN/dz$ towards
  CDQ/LDQ, give the ratios (excess factors) for strong
  \mgii absorbers as:
     
\begin{equation}
Ex = \left( \frac {dN}{dz} \right)_{\rm CDQs} / \left( \frac {dN}{dz} 
\right)_{\rm QSO} = 1.09\pm^{     0.06}_{     0.06},
\label{eq:ex_cdq}
\end{equation}

\begin{equation}
Ex = \left( \frac {dN}{dz} \right)_{\rm LDQs} / \left( \frac {dN}{dz} 
\right)_{\rm QSO} = 1.01\pm^{      0.10}_{      0.10},
\label{eq:ex_ldq}
\end{equation}

Thus, we detect only a marginally significant ($\sim 1.5\sigma$)
excess of `intervening' strong \mgii absorbers towards CDQ versus QSO
sightlines. Although in Paper I, no statistically significant excess
was detected for CDQs, this implies no real contradiction, in view of
about 35 times smaller CDQ sample available in Paper I. The same
explanation probably holds for the non-detection by
\citet{Ellison2004ApJ...615..118E} who used an even smaller sample of
CDQs.
      
 \begin{figure*}	
   \centering	
      \includegraphics[width=7.0cm]{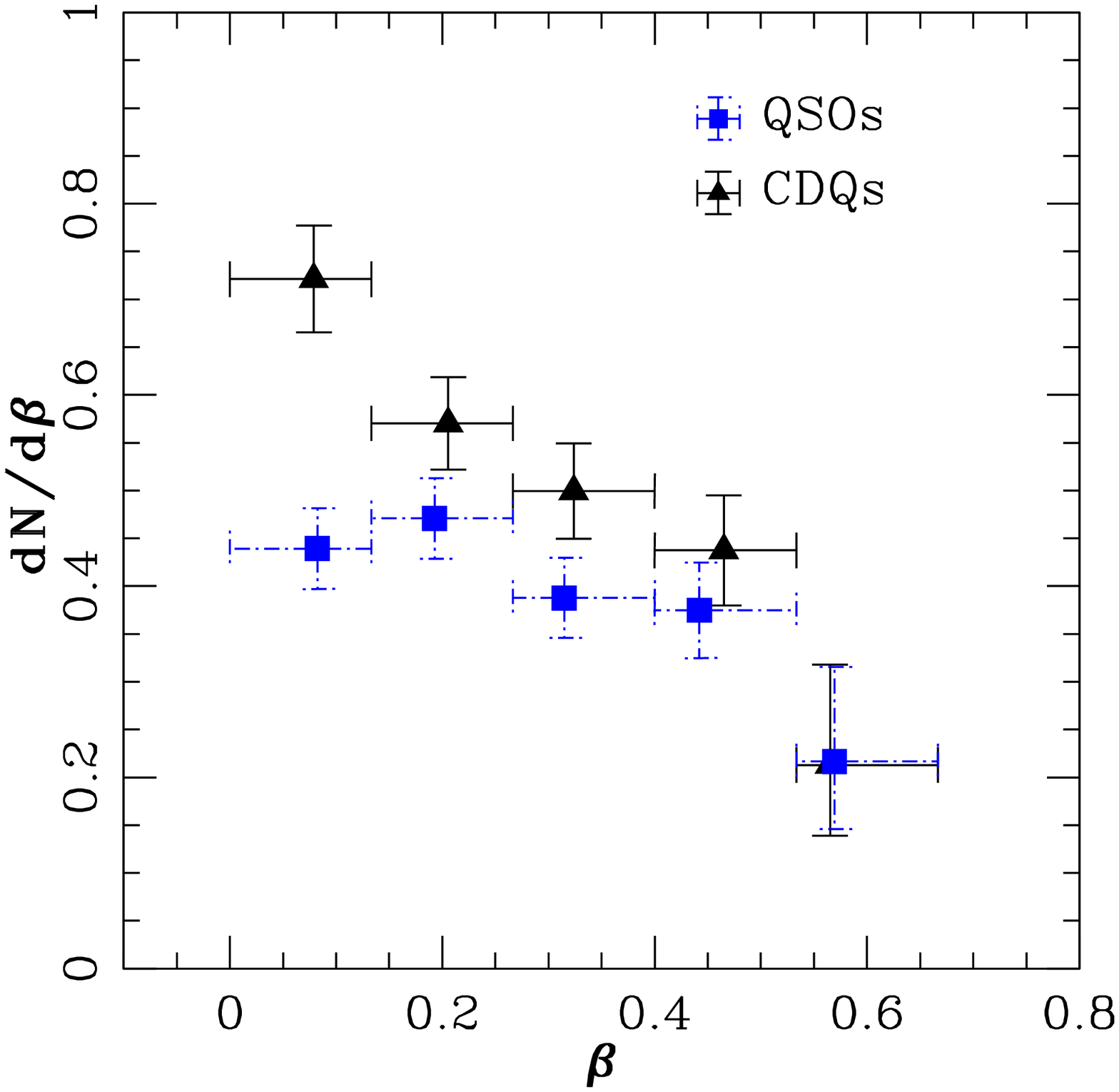}
      \includegraphics[width=7.0cm]{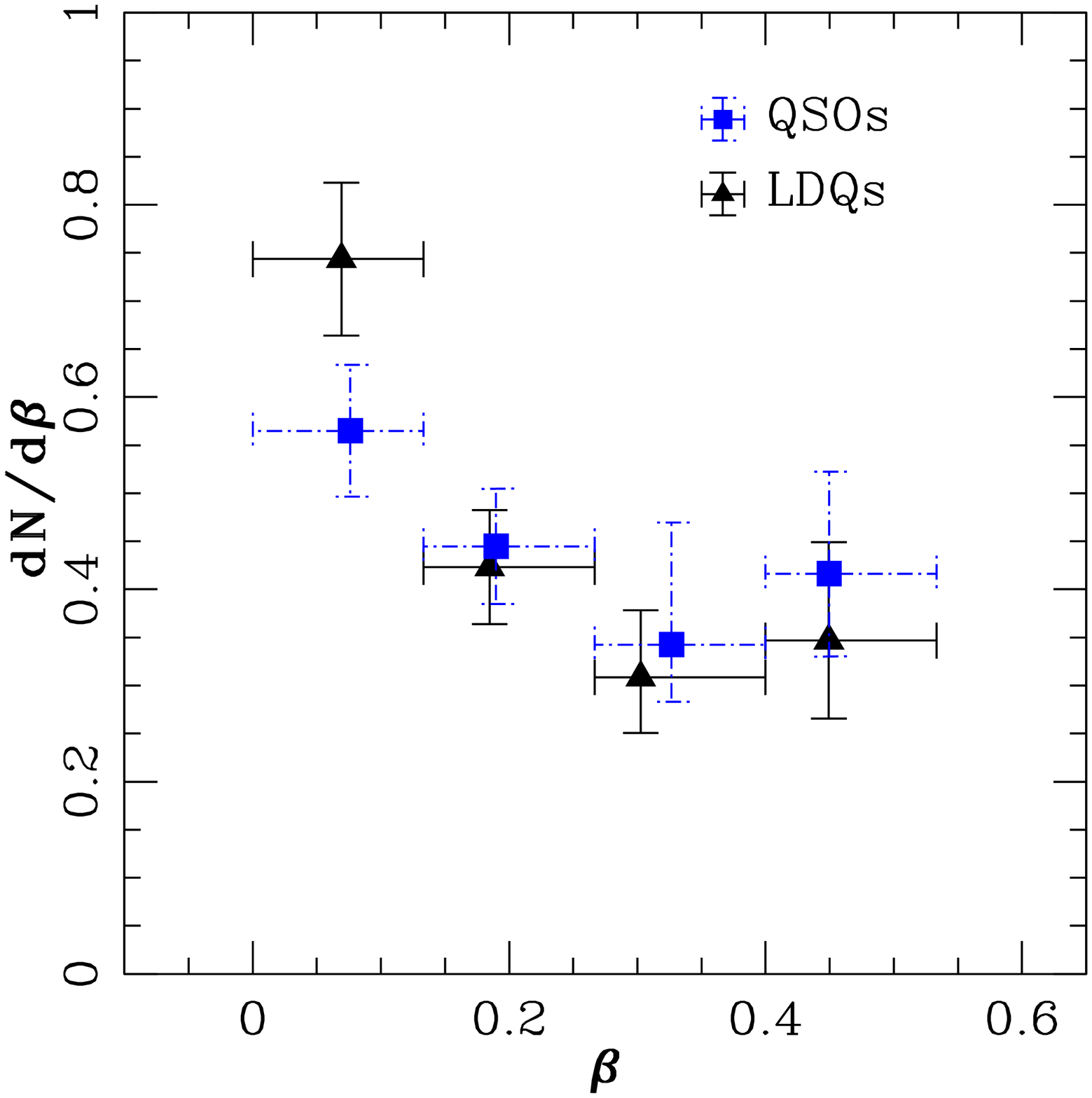}
 \caption {\emph{Left:} The incidence of \mgii strong absorbers as a function of offset
   velocity, $v=\beta c$, measured relative to the quasar rest-frame,
    for CDQs and QSOs. \emph{Right:} same for LDQs
   and QSOs}. The horizontal
   bars indicate the velocity bins and vertical bars the $1\sigma$
   uncertainty from counting statistics.
 \label{fig:beta_sdss_cor_lob_nor}
   \end{figure*}

In Figure~\ref{fig:dNdz}, we have displayed for different redshift
bins, the values of $dN/dz$ determined for `intervening' strong \mgii
absorption systems towards CDQs and then compared them with the
analytical fit taken from \citet{Zhu2013ApJ...770..130Z} for normal
QSOs. The null hypothesis that the two distributions are statistically
indistinguishable is ruled out by a \chisq-test, for the CDQ versus
QSO sightlines, at $1.6\sigma$. This is further
discussed in Section~\ref{sec:discussion_mgiidndz}.

\subsection{The distribution of offset velocity (v) for `intervening' strong \mgii absorbers}
\label{subsec:beta_intorigin}	

 To probe further the marginal excess of $dN/dz$ found here towards
 CDQs vis a vis QSOs, we now compute the distribution of offset
 velocity, $v$, of the strong \mgii absorption systems (observed at $z
 = z_{abs}$), relative to the background quasar ($z = z_{em}$), where
 $v$ is given by BBM,

\begin{equation}
\frac{v}{\rm c}  = \frac {(1+z_{\rm em})^2-(1+z_{\rm abs})^2}
{(1+z_{\rm em})^2+(1+z_{\rm abs})^2},
\end{equation}

 Note that in Paper I, the excess of strong \mgii absorption systems
 reported for blazars was concentrated towards the lower end of $v$
 distribution ($v/c \la 0.1$, see also BBM). The excess was attributed
 to acceleration of the absorbing gas clouds by the powerful blazar
 jets pointed close to our direction. For a meaningful check on this
 idea we need to derive a subset from our CDQ sample so that \zemi of
 each selected CDQ is such that its SDSS spectral coverage is
 continuous at least out to about $40000$~\kms\ blueward of its
 $z_{em}$, thus allowing the possibility of detecting any associated
 absorber masquerading as intervening absorber due to an unusually
 high offset velocity $v$, relative to the quasar. This requirement,
 together with the fact that $4000$\AA\ and $9000$\AA\ are the
 conservative lower and upper observed wavelength limits in the SDSS
 spectra, constrains us to adopt the range of \zemi between 0.62 and
 2.2. This redshift range contains 2919 of the CDQs and 1339 LDQs from
 our sample, in all showing 473 and 190 `intervening' strong \mgii
 absorbers, respectively. For comparison purpose, we have also
 constructed a subset of normal QSOs, via an emission redshift
 matching with each member of the CDQ subset, taking a redshift
 matching tolerance of $\Delta$\zem $\sim$ 0.003. In addition, we
 impose a similar constraint on luminosity matching, taking a matching
 tolerance of $\Delta log_{10}l_{(bol)}$$\sim$0.01, to account for the
 known correlation between bolometric luminosity ($l_{bol}$) and
 redshift in flux limited surveys like SDSS ~\citep[e.g., see
 ][]{Sharma2013MNRAS.431L..93S}. As these CDQ and QSO subsets are both
 derived from the SDSS, the observed wavelength window is common for
 them. This automatically ensures that the observable range of the
 relative velocity (i.e., offset velocity, $v$) detectable in the
 optical spectrum is quite similar for both subsets and an unbiased
 comparison of the $v-$distributions is thus possible for them. \par

 The incidence of absorbers as a function of velocity, $dN/d\beta$
 with $\beta=v/c$, for the CDQ$-$QSO and LDQ$-$QSO subsets matched in
 emission redshift and bolometric luminosity are shown in
 Figure~\ref{fig:beta_sdss_cor_lob_nor}. At velocity offset, $\sim
 0.1c$, an excess in the $dN/d\beta$ distribution can be seen for the
 CDQs compared to the normal quasars, while for LDQs the distribution
 is similar to that for the normal quasars. The null hypothesis that
 the two distributions are statistically indistinguishable is ruled
 out by a \chisq-test, for the CDQ versus QSO and LDQ versus QSO
 sightlines, at $3.75\sigma$, $0.78\sigma$ levels, respectively. 
   It is noteworthy that the $3.75\sigma$ excess  seen for CDQs comes primarily 
   from the low-$\beta$ range ($\beta = $0.05-0.1).

\section {Discussion}
\label{sec:discussion_mgiidndz}

The intriguing possibility of $dN/dz$ of strong \mgii absorbers being
dependent on the type of background source has been raised in recent
literature (Section~\ref{sec:intro_mgiidndz}). Here we have extended
those investigations using two samples of radio-loud quasars, one
showing a core-dominated radio morphology (3975 sightlines) and the
other consisting of lobe-dominated quasars (1583 sightlines). These
two samples are not only well defined, but are also very similar in
terms of redshift path distribution and wavelength coverage (since all
the spectra are taken from the SDSS), which renders them highly
suitable for the purpose of comparison. \par

Our analysis firstly shows that $dN/dz$ for `intervening' strong
\mgii absorption systems display a small excess of about
9\% ($1.5\sigma$) for CDQs in comparison to a redshift-matched sample
of optically selected QSOs (Section~\ref{subsec:comp_dndz_CLQ}).
Potential biases considered to explain this marginal excess are
extinction due to dust within the absorbers (which would lower the
$dN/dz$ towards QSOs) and gravitational lensing amplification of the
background optical continuum, which could enhance $dN/dz$ towards
CDQs/blazars (see Section~\ref{sec:intro_mgiidndz}). Recently, BBM
have estimated that both these phenomena are unlikely to noticeably
influence the counts of strong \mgii absorption systems towards blazar
sightlines. As a more plausible alternative, they have attributed the
observed excess of \mgii absorbers towards blazars (in comparison to
QSOs) to absorbing gas clouds swept up by the powerful blazar jets and
accelerated to mildly relativistic speeds
\citep[also,][]{Wild2008MNRAS.388..227W, Krause2002A&A...386L...1K}.
BBM have estimated that for reasonable values of jet power and ambient
gas density, a column of cool gas as large as $10^{18}-10^{20}
cm^{-2}$ can be swept up and accelerated to velocities of up to $\sim$
0.2c, if not higher. This is consistent with the result presented in
Paper I where a hint was presented for strong \mgii absorption systems
to be accelerated by blazar jets up to $v \sim$ 0.15c. \par

Compared to blazars, the jets in CDQs and LDQs, are progressively less
well aligned from the line-of-sight
\citep[e.g.][]{Antonucci2012A&AT...27..557A,
  Barthel1999ASPC..162..127B, Urry1995PASP..107..803U}. This is in
accord with the present work which has revealed a mild ($\sim$ 9\%)
excess of strong \mgii absorbers towards CDQs, relative to QSOs, and
no excess towards LDQs. More specifically, the distribution of
$dN/d\beta$ shown in Figure~\ref{fig:beta_sdss_cor_lob_nor} has
revealed for CDQs a significant excess at offset velocities around
$0.1c$, while comparing with QSO. The null hypothesis that their
$dN/d\beta$ distributions are identical is ruled out at $3.75\sigma$
level using the \chisq-test (see Section~\ref{subsec:beta_intorigin})
Thus, the results of this study and Paper I are both consistent with
the proposal that cool gas clouds capable of producing strong \mgii
absorption can be accelerated by the jets of powerful AGN and/or due
to the accretion-disk outflows, even up to mildly relativistic speeds
i.e., several times more than the canonical limit of
$5000$~\kms\ adopted in the literature for associated absorbers. This
underscores the importance of a more realistic numerical modelling of the
interaction of collimated outflows from the nuclei of quasars/blazars,
particularly their jets with the ambient medium (see also, BBM). \par

\begin{figure}	
   \centering	
      \includegraphics[width=7.0cm]{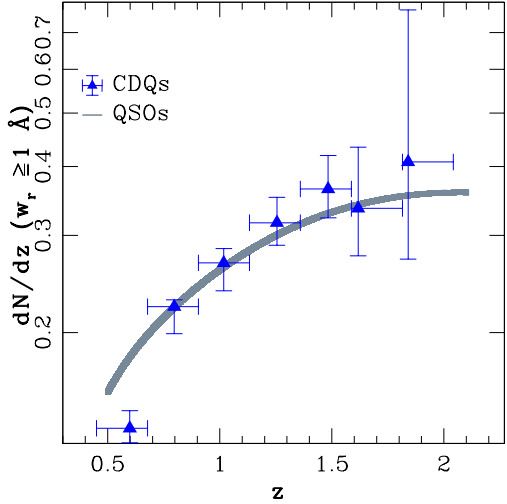}
\caption{ Number density evolution of strong ($W_{\rm r}(2796) \ge
  1.0$~\AA) \mgii absorption systems towards our  matched sets of SDSS CDQs
  (triangles) and SDSS QSOs (solid line). The solid line has been
  computed from the analytical expression given by
  \citet{Zhu2013ApJ...770..130Z} for strong \mgii absorption systems
  towards QSOs. Only the spectral regions corresponding to $v > 0.2c$
  have been included in computing  this plot (Section~\ref{subsec:beta_intorigin}).}
         \label{fig:cor_lob_sdss_evol_bp2c}	
         
   \end{figure}

Finally, we examine the implication of this inference for estimating
the cosmological evolution of the purely intervening strong \mgii
absorbers, by repeating the analysis of
Section~\ref{subsec:comp_dndz_CLQ} after excluding all redshift paths
corresponding to $v < 0.2c$. The result of this analysis are shown in
Figure~\ref{fig:cor_lob_sdss_evol_bp2c} (which is a refined version of
Figure~\ref{fig:dNdz}). The $dN/dz$ distributions for CDQs and QSOs
now appear quite similar (a \chisq-test shows that the difference is
significant only at $0.6\sigma$ level), as indeed is expected for
bona-fide intervening absorbers. In contrast, it may be recalled that
by setting the offset velocity exclusion limit at the conventional
value of $5000$~\kms, a $1.6\sigma$ excess of $dN/dz$ was found for
CDQs around the redshift range $1.2 - 1.8$ (Figure~\ref{fig:dNdz}).
This again goes to support the inference that associated strong \mgii
systems may well occur at velocity offsets that are an
order-of-magnitude larger than the conventionally adopted limit of $v=
5000$~\kms. This may have interesting physical interpretation
  such as what fraction of CDQs have cool gas outflows 
\citep[e.g., see][]{Nestor2008MNRAS.386.2055N} as outlined below.\par

 From the excess of $dN/d\beta$ seen for CDQs compared to QSOs at
  low $\beta$ values (Figure~\ref{fig:beta_sdss_cor_lob_nor}), we can
  compute the fraction of strong \mgii absorber within $\beta \le 0.1$
  (i.e., $v \le 0.1c$) that are likely to be associated with CDQs
  itself, similar to the \civ outflow fraction analysis by
  \citet[][]{Nestor2008MNRAS.386.2055N}. For this purpose we take
  intervening \mgii absorbers seen towards normal quasars within
  $\beta \le 0.1$ as the model for intervening absorbers for our CDQ
  sightlines (e.g., see Figure~\ref{fig:beta_sdss_cor_lob_nor}). Using
  this model we have then computed the excess fraction of strong \mgii
  absorbers from their observed values towards the CDQ sightlines
  (e.g., see Figure~\ref{fig:beta_sdss_cor_lob_nor}) and found this
  excess fraction to be $0.30\pm0.08$ for CDQs. We further note that
   strong \mgii absorbers were actually contributed by just
  $\sim$5\% of the sightlines out of the total CDQ sightlines we
  searched for strong \mgii absorber within $\beta \le 0.1$ range.
  Further, these strong \mgii absorber consist of associated
  absorbers, e.g., cool gas outflows (accounting for the excess
  fraction of $0.30\pm0.08$ for CDQs) plus truly intervening absorber
  (like those detected in normal QSOs). Therefore, the fraction of
  CDQs having cool gas outflowing along our direction is $\sim$ 1.5\%
  obtained by multiplying the fraction 0.05 of CDQs showing strong \mgii
  absorbers with the excess fraction $0.30\pm0.08$ of strong \mgii
  absorbers that are likely to be associated with the CDQs. \par

 Further, if we make a loose assumption that all CDQs have
  outflow, then our above result of 1.5\% sightlines having cool gas
  outflows along our direction suggest that about $\sim 25^{\circ}$
  cone angle is covered by such cool outflowing gas. However, more
  likely, the axes of CDQs are not randomly orientated but lie within,
  say $\sim 30^{\circ}$ from the line of sight
  \citep[e.g.,][]{Antonucci2012A&AT...27..557A}, as a result the above cone
  angle of the cool gas outflows decreases from $\sim25^{\circ}$ to
  $\sim 7^{\circ}$.

\section{Conclusions}
\label{sec:conclusion_mgiidndz}

Our analysis of large and well-matched samples of 1583 LDQs and 3975
CDQs, covering a redshift range 0.389$-$4.922 for the LDQs and
0.389$-$4.821 for the CDQs, has led to the following conclusions:

\begin{enumerate}

\item $dN/dz$ of `intervening' strong \mgii absorbers towards CDQs
  shows a marginal excess over the value known for normal quasars
  (QSOs), at a level of about $9\%$ with $1.5\sigma$ confidence.
  No significant excess is detected for LDQs.

\item In the redshift dependence of $dN/dz$ of strong \mgii absorbers
  towards CDQs versus normal QSOs, a $1.6\sigma$ excess is detected
  for CDQs over a narrow redshift interval $1.2 - 1.8$. However, this
  excess vanishes when we exclude the absorbers within redshift path
  corresponding to velocity offsets of up to 0.2c relative to the
  background source (see below).

\item The $dN/d\beta$ distribution for the redshift and bolometric
  luminosity matched subsets of CDQs and QSOs are found to differ,
  using the \chisq$-$test, at 3.75$\sigma$ level. This difference
  appears to be due to the excess seen in the $dN/d\beta$ distribution
  up to $v \sim0.1c$, for CDQs.

\item The noticeable excess found for CDQs at $v \la
  0.1c$, over the matched sample of QSOs, as well as the mild overall
  excess of $dN/dz$ found for CDQs, can both find a plausible
  explanation within a scenario where a significant fraction
  of the absorbers seen towards CDQs (and blazars) with offset
  velocities, $v$, of up to $\sim$0.1c might be accelerated by their
  powerful jets and/or due to the accretion-disk outflows. This
  velocity limit, initially hinted in Paper I and BBM, is several
  times the canonically adopted limit of $v = 5000$~\kms\ for
  associated absorbers.

\end{enumerate}

\section*{Acknowledgments}
 We thank an anonymous referee for the constructive 
criticism and helpful suggestions. We are also grateful to Dr. D. Lawther for 
supplying the redshift path data.\par

 G-K is supported by a NASI Senior Scientist Platinum Jubilee
 fellowship. \par

 Funding for the SDSS and SDSS-II has been provided by the Alfred P.
 Sloan Foundation, the Participating Institutions, the National
 Science Foundation, the U.S. Department of Energy, the National
 Aeronautics and Space Administration, the Japanese Monbukagakusho,
 the Max Planck Society, and the Higher Education Funding Council for
 England. The SDSS Web Site is http://www.sdss.org/. The SDSS is
 managed by the Astrophysical Research Consortium for the
 Participating Institutions. The Participating Institutions are the
 American Museum of Natural History, Astrophysical Institute Potsdam,
 University of Basel, University of Cambridge, Case Western Reserve
 University, University of Chicago, Drexel University, Fermilab, the
 Institute for Advanced Study, the Japan Participation Group, Johns
 Hopkins University, the Joint Institute for Nuclear Astrophysics, the
 Kavli Institute for Particle Astrophysics and Cosmology, the Korean
 Scientist Group, the Chinese Academy of Sciences (LAMOST), Los Alamos
 National Laboratory, the Max-Planck-Institute for Astronomy (MPIA),
 the Max-Planck-Institute for Astrophysics (MPA), New Mexico State
 University, Ohio State University, University of Pittsburgh,
 University of Portsmouth, Princeton University, the United States
 Naval Observatory, and the University of Washington. 

\label{lastpage}

\bibliography{references}
\end{document}